%% file: published.tex
\documentclass[aps,prb,twocolumn,groupedaddress,preprintnumbers,amsmath,amssymb,floatfix,10pt]{revtex4}
\usepackage{graphicx}
\usepackage[tight]{subfigure}
\usepackage{dcolumn}
\usepackage{float}
\usepackage{bm}
\usepackage{enumerate}
\usepackage[english]{babel}
\usepackage{amsfonts}
\usepackage{amsthm}
\usepackage{amssymb}
\usepackage{calc}
\usepackage{color}
\usepackage{ifthen}

\usepackage{inputenc}
\inputencoding{utf8}

\input{commands.tex}
\begin{document}

\title{Magnetic field delocalization and flux inversion in fractional
vortices in two-component superconductors 
}
\author{
Egor Babaev${}^{1,2}$,  Juha J\"aykk\"a${}^{3,4}$, Martin Speight${}^4$, }
\address{${}^1$Physics Department, University of Massachusetts, Amherst MA 01003, USA\\
${}^2$Department of Theoretical Physics, The Royal Institute of Technology, 10691 Stockholm, Sweden\\
${}^3$ Department of Physics and Astronomy, University of Turku, FI-20014 Turku, Finland\\
${}^4$ School of Mathematics, University of Leeds, Leeds LS2 9JT, UK
}
\begin{abstract}
We demonstrate that, in contrast to the single-component Abrikosov vortex, in two-component superconductors vortex solutions  with   exponentially screened  magnetic field
 exist only in exceptional cases: in the case of vortices carrying an integer number of
  flux quanta, and in a special parameter limit for half-quantum vortices. For all other parameters
  the vortex solutions have delocalized magnetic field  with a slowly decaying tail.  
Furthermore, we demonstrate a new effect which is generic in two-component systems but has 
no counterpart in single-component systems: on exactly half of the parameter space of the $U(1)\times U(1)$ Ginzburg-Landau model,  the magnetic field of a generic 
fractional vortex
inverts its direction at a certain distance from the vortex core.

\end{abstract}
\maketitle
The  two-component Ginzburg-Landau (TCGL) model, in which two
independent superconducting components interact with each 
other  via a coupling to a vector potential,
appears in various  physical contexts. 
It describes the projected quantum liquid states of metallic hydrogen and its isotopes 
under high pressure\cite{frac1,frac1b,frac2,obs}, where superconductivity
of electrons coexists with superconductivity of protons or a Bose condensate of deuterons.
Similar models describe neutron star interiors,  
where the two superconducting components represent possible protonic and
 $\Sigma^-$ hyperonic Cooper pairs \cite{Jones}. There are also various physical situations where
 the $U(1)\times U(1)$ TCGL model  arises as 
an effective description\cite{DQC}. 
The crucially important excitations appearing
 in the physics of rotational and  magnetic responses, fluctuations
and phase transitions in these systems are the topological defects (vortex lines and loops).
Qualitative analysis of the  $U(1)\times U(1)$ symmetric TCGL model
 indicates that it allows vortex excitations
carrying an arbitrary fraction of the standard
flux quantum, where the fraction is determined by a continuous parameter, the ratio 
of superfluid densities \cite{frac1}. 
There is also  growing interest in various unusual integer-flux vortex solutions which can 
be viewed, in this model, as bound states of fractional flux vortices\cite{PRB05,Moshchalkov}.
So far, fractional flux vortices in these theories
have been discussed\cite{frac1} only in the so-called London limit, a mathematical simplification wherein the condensate densities are assumed to be
constant outside the vortex core, which is modeled by a sharp cutoff.
It is well  known that in single-component systems the London limit gives a qualitatively 
accurate picture of the
behavior of the fields of a vortex in the full Ginzburg-Landau model; in particular, it correctly predicts that
the magnetic field varies monotonically and is screened exponentially at large distances.

In this Letter, we demonstrate that
 the situation in the two-component case
is entirely different. We find that vortex solutions in the TCGL  model are,  in fact, qualitatively different
from the solutions obtained in the London limit,
and exhibit highly unusual behaviour for a system which has a Meissner effect.
 Namely,  we find that the magnetic flux of a fractional vortex
is generically {\em not} exponentially localized in space, but has a { long  tail
which decays according to a $1/r^4$ power law.} 
The magnetic field has a tendency to get extremely
delocalized for small fractions of flux quanta, where the maximum 
of the magnetic field at the vortex center becomes barely distinguishable.
This effect can be understood using
explicit asymptotic formulas we obtain 
for the magnetic field and condensate densities at long range in terms 
of the TCGL model parameters. These formulas show, moreover, that under quite generic conditions
in multicomponent superconductor the  magnetic field in a fractional flux vortex  can {\it reverse} its direction
at a certain distance from the core, in stark contrast to vortex solutions in 
single-component superconductors. 

The system of interest is the $U(1)\times U(1)$ symmetric TCGL model with free energy
\bea
&&E=\frac12\int dx\,
dy\left\{|(\cd_k+ieA_k)\psi_1|^2+|(\cd_k+ieA_k)\psi_2|^2 
\right.\nonumber\\
&&+\eta_1(u_1^2-|\psi_1|^2)^2+\eta_2(u_2^2-|\psi_2|^2)^2+ \left. (\epsilon_{ij}\partial_i A_j)^2 \right\}.
\label{model}
\eea
Here $\psi_{1,2}$
are two complex scalar fields corresponding to two superconducting order 
parameters. 
The model (\ref{model}) is realized in  physical 
systems where the electrodynamics is local and 
Josephson-like 
coupling between condensates is forbidden. We have given the condensates equal electric charge, but
the results apply equally well for a system of oppositely charged condensates\cite{frac1b,frac2,obs,Jones} since the model (\ref{model}) is invariant 
under inversion of the sign of the charge of a condensate  
 accompanied by complex conjugation of that condensate.
The results can be 
straightforwardly 
generalized to include other terms in the effective potential, or  
mixed gradient terms, so long as these are consistent  with the
$U(1)\times U(1)$ symmetry.
Vortices in this model are solutions of the Euler-Lagrange equations
\bea
(\cd_k+ieA_k)^2\psi_1+2\eta_1(u_1^2-|\psi_1|^2)\psi_1&=&0\label{f1}\\
(\cd_k+ieA_k)^2\psi_2+2\eta_2(u_2^2-|\psi_2|^2)\psi_2&=&0\label{f2}\\
-\epsilon_{kj}\cd_jB&=&J_k\label{f3}
\eea
where $J_k$ is the  supercurrent
$J_k=\frac{i}{2}e\{\psi_1{(\cd_k-ieA_k)\psi_1^*}-{\psi_1^*}(\cd_k+ieA_k)\psi_1 
+\psi_2{(\cd_k-ieA_k)\psi_2^*}-{\psi_2^*}(\cd_k+ieA_k)\psi_2\}.$
In the first part of this paper we seek solutions of this system within the axially symmetric 
ansatz
\beq
(A_1,A_2)=\frac{a(r)}{r}(-\sin\theta,\cos\theta);\ \psi_i=\sigma_i(r)e^{-in_i\theta} \label{a1}
\eeq
which amounts to imposing $2\pi n_i$ winding on the phase of condensate $\psi_i$, where
$n_i$ are integers.
Here $x+iy=re^{i\theta}$ and $a(r),\sigma_i(r)$ are real 
profile functions.
Note that,  in certain cases the axial symmetry of vortex solutions in this
model
was found to be spontaneously broken \cite{PRB05}. However in the cases
studied below, the solutions are axially symmetric, as
 verified by the numerical simulations presented in the second part of the 
paper. 
 In what follows we are looking for localized solutions  in the sense
that $|\Jv|\ra 0$ and $\sigma_i\ra u_i$ as $r\ra\infty$.
It follows
that
\beq
a(r)\ra a_\infty=\frac{1}{e}\Phi,\quad \mbox{where}\quad
 \Phi=\frac{n_1u_1^2+n_2u_2^2}{u_1^2+u_2^2}.
\eeq
By Stokes's theorem, it follows that the total magnetic flux through the
$xy$ plane is $\int B\, dx\, dy = \frac{2\pi}{e}\Phi$
which is a fractional multiple $\Phi$
of the usual flux quantum if $n_1\neq n_2$ \cite{frac1}.

Substituting (\ref{a1}) 
 into 
(\ref{f1}),(\ref{f2}),(\ref{f3}) yields a coupled system of 
ordinary differential equations
\bea
\sigma_1''+\frac{\sigma_1'}{r}-\frac{(n_1-ea)^2}{r^2}\sigma_1+2\eta_1(u_1^2-\sigma_1^2)\sigma_1&=&0\label{o1}\\
\sigma_2''+\frac{\sigma_2'}{r}-\frac{(n_2-ea)^2}{r^2}\sigma_2+2\eta_2(u_2^2-\sigma_2^2)\sigma_2&=&0\label{o2}\\
a''-\frac{a'}{r}-e(ae(\sigma_1^2+\sigma_2^2)-n_1\sigma_1^2-n_2\sigma_2^2)&=&0\label{o3}
\eea
subject to the
boundary conditions $a\ra a_\infty$, $\sigma_1\ra u_1$, $\sigma_2\ra u_2$
as $r\ra\infty$. 
 Solutions with $n_1=n_2$ carry integer flux and 
 were considered in Ref.\cite{PRB05}.  
They turn out to have a much richer
variety of interaction behavior than Abrikosov vortices. However, just like 
their single-component counterparts, 
the modulation of the fields $|\psi_i|$ and
$|B|$ is exponentially  localized in space. Here we observe that, by contrast,
 if $n_1\neq n_2$,
neither $n_1-ea$ nor $n_2-ea$ approaches zero as $r\ra\infty$, and
consequently   it follows  from
(\ref{o1}), (\ref{o2}) that neither $\sigma_1$ nor $\sigma_2$ can 
approach its boundary value ($u_1,u_2$ respectively) exponentially fast.
So, in contrast to integer flux vortices,  for fractional flux vortices 
the densities $|\psi_i|$ can recover their asymptotic values only 
according to some  power law.
 Since the third terms in (\ref{o1}), (\ref{o2})
decay like $r^{-2}$ it is consistent to assume (the assumption is verified 
below) that
\beq\label{sig}
\sigma_i(r)\sim u_i-\alpha_ir^{-2},\quad i=1,2
\eeq
at large $r$, for some real coefficients $\alpha_1,\alpha_2$. Then
$\sigma_i''$, $\sigma_i'/r$ are $O(r^{-4})$, and demanding that the leading
term (order $r^{-2}$) vanishes gives the prediction
\beq\label{alpha}
\alpha_i=\frac{(n_i-\Phi)^2}{4\eta_iu_i},\quad i=1,2.
\eeq
Note that $\alpha_i>0$, so $\sigma_i$ approaches its boundary value from below,
as one expects.
From (\ref{o3}), it is then consistent to assume (again, verified below)  that
\beq\label{b5}
a(r)\sim\frac{\Phi}{e}-\beta r^{-2}
\eeq
at large $r$, for some real coefficient $\beta$. Again, $a'$, $a'/r$ are
order $r^{-4}$, and demanding that the leading term in (\ref{o3}) vanishes
leads one to predict that
\beq
\label{beta}
\beta=\frac{1}{2e(u_1^2+u_2^2)}\left\{
\frac{(n_1-\Phi)^3}{\eta_1}+\frac{(n_2-\Phi)^3}{\eta_2}\right\}.
\eeq
Now $B=r^{-1}a'(r)$, so in the case where $\Phi>0$ (e.g.\ if
$n_1,n_2\geq 0$), $a(r)$ interpolates between $a(0)=0$ and $a_\infty>0$,
so one expects $a'(r)>0$ uniformly, and hence $B(r)>0$. In particular,
one expects $a(r)$ to approach its boundary value $a_\infty$ from below, 
so that $\beta>0$. But in this regard, formula (\ref{beta})
contains a surprise: it is quite possible for $\beta$ to be negative. In this
case, since $B(r)\sim 2\beta r^{-4}$ at large $r$, we see that {\it the
magnetic field has to flip its direction as one travels out from the vortex core:
it is positive for small $r$ and negative for large $r$}.
Let us introduce polar
coordinates on the $u_1u_2$ and $\eta_1\eta_2$ parameter planes, so
$u_1+iu_2=ue^{i\zeta}$ and $\eta_1+i\eta_2=\eta e^{i\phi}$
where $0<\zeta,\phi<\frac\pi2$. Then
\beq
\beta=\frac{(n_1-n_2)^3}{2eu^2\eta}\left\{\frac{\sin^6\zeta}{\cos\phi}-
\frac{\cos^6\zeta}{\sin\phi}\right\},
\eeq
so $\beta<0$ if and only if $\tan\phi<\cot^6\zeta$, which holds on
precisely half of the $\zeta\phi$ square. 
 Hence, 
not only can magnetic flux reversal occur
for fractional flux vortices, it is a {\it generic} effect which occurs on half the
parameter space of the TCGL model (see also remark \cite{remark1}).

It is interesting to consider parameter values on the curve
$\tan\phi=\cot^6\zeta$, for which $\beta\equiv 0$. At generic points
on this curve, $a(r)\sim a_\infty-\beta'r^{-4}$, so 
for that family of vortices  the magnetic field $B$ is 
power-law localized, but with unusual power, decaying as $r^{-6}$. However we find that a very 
special
situation  happens when the vortex  carries
a half of the flux quantum and both condensates have the same coherence 
length, that is,  $u_1=u_2$, $\eta_1=\eta_2$ (i.e.\ $\phi=\zeta=\frac\pi4$).
This regime is relevant for physical situations where such a TCGL
model is dictated by symmetry.
Substituting power series ans\"atze 
$\sigma_i(r)=\sum_{k=0}^\infty\alpha_{i,k}r^{-k},\quad
a(r)=\sum_{k=0}^\infty\beta_kr^{-k}$
into (\ref{o1})-(\ref{o3}), we see that it is consistent that
$a(r)=a_\infty$ to all orders (i.e.\, $\beta_k=0$ for $k\geq 1$): equations
(\ref{o1}) and (\ref{o2}) then imply that $\sigma_1(r)=\sigma_2(r)$ to all
orders (i.e.\ $\alpha_{1,k}=\alpha_{2,k}$ for all $k$), which is
consistent with (\ref{o3}) (whose left hand side is then zero to all orders). 
One
is led to conclude, therefore, that exponential localization of  
the magnetic field $B(r)$
is recovered for the half-quantum vortex at this symmetric parameter set, 
despite the density fields $|\psi_i (r)|$ still being only $r^{-2}$ 
localized.

To obtain more detailed knowledge of the behavior of the profile
functions of fractional flux vortices, and confirm accurately the above 
calculations,
we must perform   numerical computations.
The shooting method for system (\ref{o1})-(\ref{o3})
described in Ref.\cite{PRB05} turns out to be hopelessly
unstable for fractional flux vortices, so we must resort to a relaxation method.
We have discretized the system using 
the method described in \cite{Jaykka}
then used
gradient based optimization algorithms to find highly accurate minima of the 
system 
energy for a given  phase winding. 
It should be noted that the numerical scheme does not impose
rotational symmetry, so if we obtain axially symmetric solutions (as we do),
we can be confident that they are stable against all small perturbations.

In this second part of the paper,
 we present the numerical results for the parameters  $e=2$, $\eta_1 = \eta_2 =
2/3$ and  several values of $u_i$. These parameters allow us to 
confirm  numerically the analytic calculations from the first part of the paper.
The characteristic unusual features 
become  more pronounced with decreasing $e,\eta_i$ (i.e. the vortex solution gets 
more delocalized and has more pronounced field inversion tail). However our 
choice of  parameters here   is motivated by minimizing the  effects of the 
boundary of the numerical grid.
We present detailed numerical investigations
of the following cases: $[n_1=1,n_2=1,u_1=1,u_2=\sqrt{0.2}]$ (flux fraction $\Phi=1$),
$[n_1=1,n_2=0,u_1=1,u_2=\sqrt{0.2}]$  (flux fraction $\Phi=5/6$)
and $[n_1=1,n_2=0,u_1=\sqrt{0.2},u_2=1]$   (flux fraction $\Phi=1/6$).

\begin{figure}[h]
  \centering
  \includegraphics[width=0.95\columnwidth]{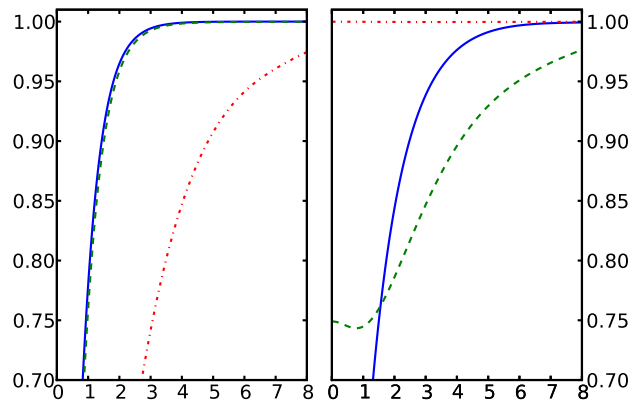} \caption{(Color online.) Asymptotic behavior of the fields in the two-component
    vortex: $|\psi_1|$ (left) and $|\psi_2|$ (right) with flux fractions $1$ (solid blue), $5/6$ (dashed green) and $1/6$
    (dash-dot red). In accord with analytic calculations, in the case of $1/6$ flux quantum, $|\psi_1|$ 
 is strikingly delocalized; however in the case of $5/6$ flux quantum, the power-law tail is tiny
    and  the difference from the integer-flux case is barely visible. The $\psi_2$ configuration is
    coreless, but  has a dip and local maximum at the origin. The dip  is especially pronounced in the case of $5/6$ flux
    quanta, and is almost invisible in the case of $1/6$ flux quanta (where $|\psi_2(0)| = 1$ and $|\psi_2(3.2)|\approx 0.9996$). }
  \label{fig:psi1psi2fig}
\end{figure}

\begin{figure}[h]
  \centering
  \includegraphics[width=0.95\columnwidth]{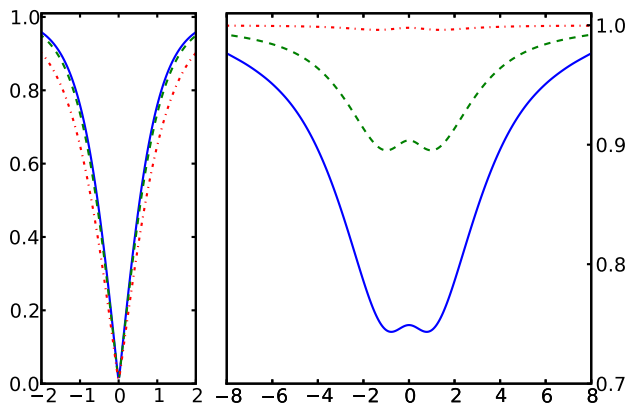}
  \caption{(Color online.) 
The behavior near the vortex core: $|\psi_1|$ (left) and $|\psi_2|$ (right) with flux fractions $5/6$
    (solid blue), $5/7$ (dashed green) and $1/3$ (dash-dot red). The component with the phase winding $|\psi_1|$ always has a
    singularity. The other component always has a non-singular ``$W$'-shaped suppression of density.}
  \label{fig:psi1psi2dip}
\end{figure}

Eq.\ (\ref{sig}) indicates that the rate at which the density
 approaches
its ground state value at large distances decreases as its corresponding 
$\alpha_i$ increases. This is indeed
confirmed by the plots in Fig.~\ref{fig:psi1psi2fig}. The long distance behavior 
of all of these agrees with (\ref{alpha}): in the integer-flux case the densities recover their vacuum values
exponentially fast, as in the case of the Abrikosov vortex 
(and $\alpha_1=\alpha_2=0$)
while in the fractional-flux cases the behavior is  $\alpha_i/r^2$. 
We also find that the component  $\psi_2$ (which does not have phase winding)
exhibits very unusual behavior near the origin in the second case: 
its density
has a local
 maximum in the core. Observe that in our model we do not 
have terms in the effective potential corresponding to direct 
interspecies density-density interactions,
and this unusual density maximum in the core is caused purely by 
electromagnetic interaction of the condensates.
We explored this behavior for a range of different values of $u_2$.
The results
of three characteristic cases with $u_2 \in \{0.2,0.4,2\}$ are shown in Fig.~\ref{fig:psi1psi2dip}, which suggests that
decreasing $u_2$ deepens the $``W"$-shaped density modulation  in the condensate
without phase winding. The maximum of $\psi_2$  originates in the fact that the circulation of 
the supercurrent in the component $\psi_2$ stems from the vector potential (see eq. (\ref{f2})).
At distances $r \ll \lambda$ from the core we have
$\sigma_1 \sim  r^{n_1}$,  $\sigma_2 \sim r^{n_2}$, $a \sim  r^2$.
The behavior of $a$ shows that
there is almost no supercurrent circulation in $\psi_2$ near the origin of the vortex.
Consequently $|\psi_2|$ tries to minimize the energy by
 recovering the  ground state value of density at short $r$. 
Since there are no singularities of superfluid velocity in the component $\psi_2$
the ``$W$''-shaped density suppression can be arbitrarily deep; however it can
never produce a zero-density  singularity in $|\psi_2|$.

\begin{figure}[h]
  \centering
  \includegraphics[width=0.95\columnwidth]{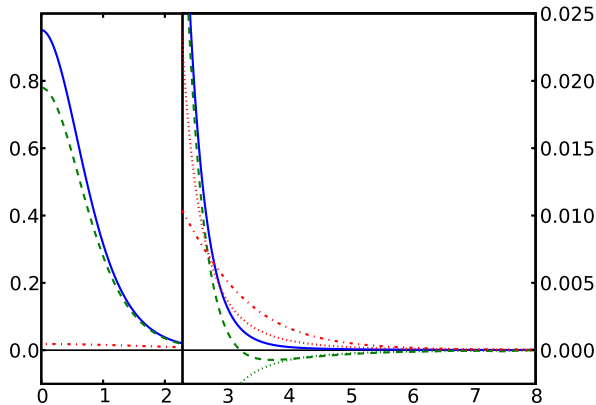}
  \caption{(Color online.) The behavior of $B_z$ near the origin of the vortex (left panel) and long-range tail
    (zoomed in, right panel) with flux fractions $1$ (solid blue), $5/6$ (dashed green) and $1/6$ (dash-dot red). We see behavior
    strikingly different from the Abrikosov vortex: in case of $1/6$-quantum vortex, the magnetic
    field is extremely delocalized without a pronounced maximum at the origin, but already at $r\approx 3.5$
    having larger value than the field of the one-quantum vortex . In case of $5/6$ quantum, the vortex accumulates magnetic
    flux {\it larger than $(5/6)\Phi_0$} near the origin, almost mimicking in this region the Abrikosov vortex.
    However the magnetic field rapidly goes to zero at $r=3.275 \pm 0.0125$, after which point the magnetic field
    flips its direction, producing a slowly decaying power-law tail of inverse flux. The delocalized magnetic flux in the
    outer region subtracts from the strongly localized flux near the origin to produce net flux $(5/6) \Phi_0$ .
The dotted lines in the right panel depict the curves predicted by Eq. \eqref{b5}.}
  \label{fig:B_whole}
\end{figure}

Let us turn our attention to the magnetic field. From the above analytic 
considerations, we expect the magnetic field to approach zero
exponentially if the flux fraction is an integer. Also exponential  and high algebraic power $1/r^6$ localization of 
magnetic field is found in some cases for half-quantum vortices. But in the general case, the magnetic field 
should have $1/r^4$ asymptotic behavior.
 Indeed, this
can be seen in Fig.~\ref{fig:B_whole}, which shows the 
magnetic field behavior in
the same three cases whose  density plots appear in Fig.~\ref{fig:psi1psi2fig}.
Figure (\ref{fig:B_whole}) confirms  the
two main generic features of vortex solutions in the TCGL model predicted in the first  part of the paper:
the delocalization of magnetic flux
when the fraction of the flux quantum is 1/6,
and the delocalization and reversal of magnetic flux when
the fraction of the flux quantum is 5/6. These features get even more pronounced
for weaker potentials and larger penetration lengths.

In conclusion, we showed that, 
quite counter-intuitively, considering the solutions of the complete
two-component Ginzburg-Landau problem  
reveals new and unusual physics.
Namely, we find that 
for generic fractional flux vortex solutions
(except for  the special parameter set of half-quantum vortices)
 the  magnetic field is delocalized, possessing 
a slowly decaying $1/r^4$ tail, and that 
on exactly half of the model's parameter space, the vortices
 exhibit magnetic flux inversion:
near the origin of the vortex there is a peak in magnetic field carrying  flux
in the positive direction of the z-axis, while at a certain distance from the core this field has a rapid reversal of 
direction
 producing a 
tail of magnetic field  in the negative direction along the z-axis.
These phenomena should have a number of physical consequences.
Field delocalization and inversion can serve as an experimental signature of  fractional vortices in 
superconductors with multiple components
or in artificial superconducting structures with several
magnetically coupled superconducting components.
The model  describes the projected quantum fluid 
of metallic hydrogen \cite{frac2,frac1b,obs}, a subject of renewed experimental pursuit.
This magnetic field delocalization effect should affect
magnetic-response-based techniques proposed to be the main tool 
to detect the transition to the quantum fluid 
of metallic hydrogen and  suggested similar transitions in  hydrogen-rich alloys 
and deuterium \cite{obs}.

JJ was supported by the Academy of Finland (Project No. 123311) and the UK
Engineering and Physical Sciences Research Council.

\end{document}

%% file: commands.tex
\theoremstyle{definition}
\ifthenelse{\not\isundefined{\thechapter}}{

}{}

\newtheorem*{huom*}{Huomautus}
\newtheorem*{seuraus*}{Seuraus}
\newtheorem*{esim*}{Esimerkki}

\newcommand{\beq}{\begin{equation}} 
\newcommand{\eeq}{\end{equation}} 
\newcommand{\bea}{\begin{eqnarray}} 
\newcommand{\eea}{\end{eqnarray}} 
\newcommand{\ben}{\begin{eqnarray*}} 
\newcommand{\een}{\end{eqnarray*}} 
\newcommand{\ra}{\rightarrow}

\newcommand{\cd}{\partial}

\newcommand{\Jv}{{\bf J}} 

%% file: published.bbl
\begin{thebibliography}{99}
\bibitem{remark1} 
From the given asymptotical analysis it follows that these effects occur
in $U(1)\times U(1)$ symmetric  models when 
the condensates have differing phase winding, and
 recover their ground state values at different rates.
For this reason the effects should also be present if the effective potential
includes density-density interaction terms such as
 $|\psi_1|^2|\psi_2|^2$
because (i) these preserve $U(1)\times U(1)$ symmetry and (ii) in their
presence the two condensates generically have distinct recovery rates.
\end{thebibliography}
